\documentclass[10pt,twocolumn,english,aps,prl,reprint,
preprintnumbers,amsmath,amssymb,colorlinks=true,linkcolor=blue,
citecolor=blue,urlcolor=blue]{revtex4-1}
\usepackage[T1]{fontenc}
\usepackage[latin9]{inputenc}
\usepackage{array}
\usepackage{float}
\usepackage{textcomp}
\usepackage{multirow}
\usepackage{amsmath}
\usepackage{graphicx}

\makeatletter

\usepackage{graphicx,epsfig}
\usepackage{longtable}
\usepackage{xcolor}

\usepackage[caption=false]{subfig}

\makeatother

\usepackage{babel}

\begin{document}

\title{No-flux boundaries stabilize scroll rings in excitable media with negative filament tension}


\author{Arash Azhand\textonesuperior{}, 
 Rico Buchholz\texttwosuperior{},
 Florian Buchholz\textonesuperior{},
Jan F. Totz\textonesuperior{}, Harald Engel\textonesuperior{}}
\email{azhand@itp.tu-berlin.de}
\address{\textonesuperior{}Institut für Theoretische Physik, Technische Universität
Berlin, Hardenbergstrasse 36, D-10623 Berlin, Germany\\
\texttwosuperior{}Theoretische Physik V, Universität Bayreuth, Univeristätsstrasse
30, D-95440 Bayreuth, Germany}

\begin{abstract}
Scroll rings in an unbounded excitable medium with negative line tension undergo an instability
ending eventually in a ``turbulent'' state, known as scroll wave (Winfree) turbulence. In this 
paper we demonstrate by numerical simulations based upon the Oregonator model for the photosensitive
Belousov-Zhabotinskii reaction (PBZR) that the Winfree turbulence is suppressed by the interaction of 
the scroll ring with a confining Neumann boundary. Instead of the Winfree turbulence a stable scroll ring
forms due to the boundary interaction. Furthermore, we will discuss the conditions under which boundary-stabilized
scroll rings could be observed in the PBZR, taking into account a light-induced excitability gradient in 
parallel to the scroll ring's symmetry axis.
\end{abstract}

\pacs{Valid PACS appear here}

\maketitle


\section{Introduction}

Excitable media can be found in many parts of biology and physics.
Under non-equilibrium conditions pattern formation can appear in many
of those systems. A common pattern that may develop is the spiral
wave. It has been observed in slime molds \cite{Siegert:PhysicaD:91},
on platinum surfaces \cite{Jakubith:PhysRevLett:90}, chemical systems
\cite{Winfree:Science:72} and in the heart \cite{Davidenko:Nature:92}.

Some parts of the human heart tissue, especially at the ventricles,
are thick enough to support not only spirals, but also three dimensional
structures, for example scroll waves and scroll rings. This
makes an investigation of these structures important, too.

The heart tissue can provide three dimensional structures, but the
boundaries are never too far away. Because of this not only the evolution
of spirals and scrolls in an unbounded domain need to be known but
also the interaction with the boundary. For spirals the interaction
with straight and circular boundaries is well studied \cite{Aranson:PhysRevE:94,Gomez-Gesteira:PhysRevE:96,Eguiluz:IntJBifurcationChaos:99,Bar:PhysRevE:03a}.
The interaction of scroll waves and scroll rings with one or more
no-flux boundaries is less discovered.

The first experimental observation of scroll rings was reported by
Winfree \cite{Winfree:Science:73} in the framework of the Belousov-Zhabotinskii
reaction. Later scroll rings were also observed in fibrilating cardiac
tissue \cite{MedvinskyPanfilovPertsov1984}.

One of the most important questions here is dealing with the time-evolution
of scroll rings, that is whether the radius of the ring is shrinking
(``positive filament tension'') or expanding (``negative filament
tension'') in time. Several experimental \cite{Winfree:SciAm:74,Welsh:Nature:1983,WinfreeJahnke1989,Bansagi:PhysRevLett:2006}
and theoretical \cite{PanfilovPertsov1984,Keener:PhysicaD:88,Biktashev:PhilTransRSocA:94}
studies revealed the dynamics of contracting scroll rings.

The primary numerical prediction of scroll rings with negative filament
tension was reported by Panfilov and Rudenko \cite{PanfilovRudenko1987}.
In experiments scroll rings with negative filament tension were found
by B\'ans\'agi and Steinbock \cite{Bansagi:PhysRevE:07}, who where
also able to determine the tension of the filament. Later a connection
to negative tension instability (Winfree turbulence) was drawn by
Alonso et al. \cite{Alonso:Science:03a} and Zaritski and colleagues
\cite{Zaritski:PhysRevLett:04}. Alonso and colleagues were also able
to suppress this turbulent state by applying periodic forcing \cite{Alonso:Science:03a,Alonso:Chaos:2003}.

\begin{figure}
\begin{centering}
\includegraphics[width=0.95\columnwidth]{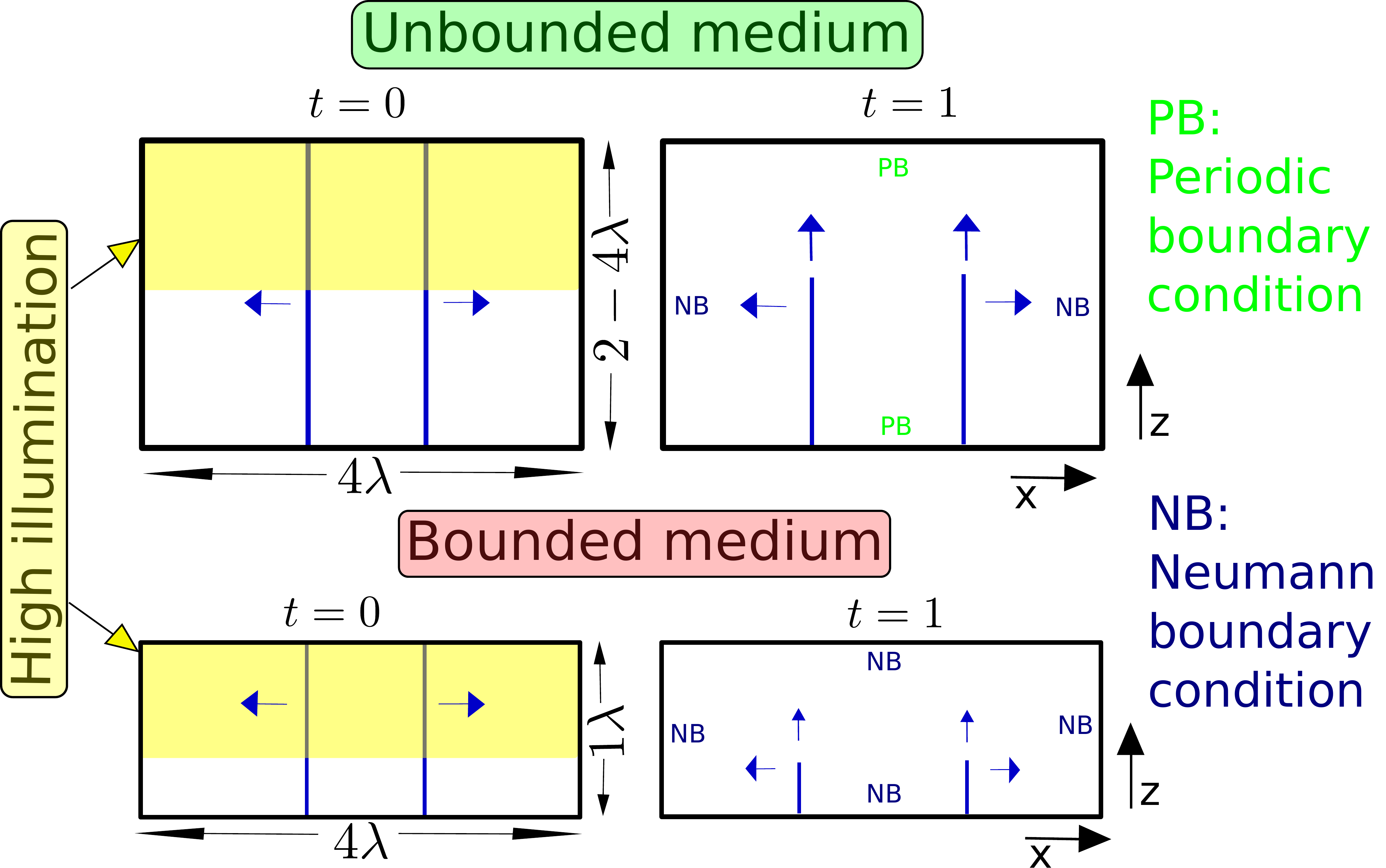}
\par\end{centering}

\caption{Schematic illustration for numerical initiation of scroll rings in
unbounded and bounded media.\label{fig:SimSpaceConditions}}
\end{figure}

Investigation of scroll rings in excitable media with inhomogeneities
were accomplished by application of parameter gradients, such as gradients
of light in the photosensitive BZ reaction \cite{Amemiya:Chaos:98}
or by applying gradients of temperature in the Ferroin catalyzed BZR
\cite{Vinson:Nature:97,Vinson:PhysRevE:99}.

Nandapurkar and Winfree \cite{Nandapurkar:PhysicaD:89} for the first
time found a stable scroll ring at a no-flux boundary through numerical investigations
of the FitzHugh-Nagumo (FHN) model in the negative filament tension
parameter regime. 

Bray and Wikswo \cite{Bray:PhysRevLett:03} studied the interaction
of two contracting scroll rings in a symmetric configuration, which
is equivalent to the situation of one scroll ring that interacts with
a Neumann boundary. 

The present work is now devoted to a systematic numerical study of
boundary-stabilized scroll rings with negative filament tension, both in
homogeneous and non-homogeneous excitable media.

The structure of this paper is as follows. First we introduce in section
\ref{sec:Model} the model that we used for the numerical simulations.
A brief summary of scroll ring filament dynamics in unbounded media
is given in section \ref{sec:Scroll-ring-unbounded}. In the subsequent
section (sec. \ref{sec:Scroll-ring-bounded}) we show how these dynamics 
change due to the interaction with a Neumann boundary and then address 
the question if the scroll rings remain stable for reasonable timescales 
(more than $100$ rotation periods of corresponding spiral waves). 
In Section \ref{sec:Towards_Exp_Verification} we will show that observation 
of boundary-stabilized expanding scroll rings in the framework of the 
photosensitive Belousov-Zhabotinskii reaction (PBZR) may be possible. 
This includes an investigation of scroll ring initiation under variation 
of media thickness versus intensity of photoinhibition (subsection 
\ref{sub:Height_Vs_Lightintensity}), evolution of boundary-stabilized 
scroll rings in media with different degrees of inhomogeneity (subsection 
\ref{sub:Inhomogeneous}), and stability of the scroll ring when 
initiated with different degrees of inclination (subsection 
\ref{sub:Inclined-scroll-rings}). Finally, section \ref{sec:Conclusion} 
is devoted to a short conclusion.

\section{Model\label{sec:Model}}

In order to simulate realistic behaviour of scroll rings in thin layers
of photosensitive BZ media, we used the \textquotedbl{}modified complete
Oregonator model\textquotedbl{} \cite{Krug:JPhysChem:90,Kadar:JPhysChemA:97a},

\begin{align}
\frac{\partial u}{\partial t} & =\frac{1}{\epsilon_{u}}\left(u-u^{2}+w(q-u)\right)+D_{u}\Delta u\nonumber \\
\frac{\partial v}{\partial t} & =u-v\label{eq:MCO}\\
\frac{\partial w}{\partial t} & =\frac{1}{\epsilon_{w}}\left(\phi+fv-w(q+u)\right)+D_{w}\Delta w,\nonumber 
\end{align}

where $u$, $v$ and $w$ are proportional to concentrations of $\text{HBr\ensuremath{O_{2}}}$
(activator), $\text{Ru\ensuremath{\left(bpy\right)_{3}^{3+}}}$ (oxidized
form of the catalyst) and $\text{B\ensuremath{r^{-}}}$ (inhibitor),
respectively. $\Delta$ is the Laplacian operator, diffusion coefficients
will be chosen as $D_{u}=1.0$ for activator $u$, $D_{w}=1.12$ for
inhibitor $w$, while there will be no diffusion for the catalyst
$v$ since in experiments the catalyst $Ru\left(bpy\right)_{3}^{3+}$
is immobilized in a gel layer. 

Parameters of the model are given by the recipe-dependent time scales
$\epsilon_{u}$ and $\epsilon_{w}$ with $\epsilon_{u}\gg\epsilon_{w}$,
ratio of rate constants $q$, stochiometric parameter $f$, and photochemically
induced bromide flow $\phi$ that is assumed to be proportional to
applied light intensity. 

We have chosen parameter sets that are shown in table \ref{tab:ParamSets}.

\begin{table}[b]
\caption{Parameter sets one and two that were investigated numerically in this 
publication. \label{tab:ParamSets}}
\begin{ruledtabular}
\begin{tabular}{ccc}
\multicolumn{1}{c}{\textrm{Parameter set}}&
\multicolumn{1}{c}{\textrm{1}}&
\multicolumn{1}{c}{\textrm{2}}\\
\hline
$\epsilon_{u}$ & $0.07$ & $0.07$ \\
$\epsilon_{w}$ & $\epsilon_{u}/90$ & $\epsilon_{u}/90$ \\
$f$ & $1.4$ & $1.16$ \\
$q$ & $0.002$ & $0.002$ \\
$\phi$ & $0.023$ & $0.014$
\end{tabular}
\end{ruledtabular}
\end{table}

A two dimensional spiral has a period of $T=6.8\,\mbox{t.u.}$
and a wavelength of $\lambda=22.6\,\mbox{s.u.}$ for the parameter
set 1 while for set 2 spirals exhibit a period of $T=6.9\,\mbox{t.u.}$
and a wavelength of $\lambda=19.6\,\mbox{s.u.}$

In both parameter sets spiral waves perform rigidly rotating spiral
tip motion. While parameter set 1 shows a transition of spiral tip
motion from outward meander via inward meander to rigid rotation as
one varies $\phi$ from lower to higher values, in parameter set 2
one observes only rigid rotation, no matter which value of $\phi$
is chosen.

Simulations where conducted with an Euler scheme for time integration
and a nineteen point star discretization of the laplacian. For space
and time discretization we used $dx=dy=dz=0.3$ and $dt=0.0005$,
respectively. Conditions for spatial geometry were always chosen such
that unbounded scroll rings could evolve in 'boxes' of $4\times4\times2\lambda^{3}$
while confined scroll rings were placed into 'boxes' with $4\times4\times1\lambda^{3}$
(see Fig. \ref{fig:SimSpaceConditions}). Neumann boundary conditions
were chosen for all sides of the medium in case of bounded media while
for the top and bottom boundaries periodic boundary conditions were
used for unbounded media.

\begin{figure}
\includegraphics[width=1\columnwidth]{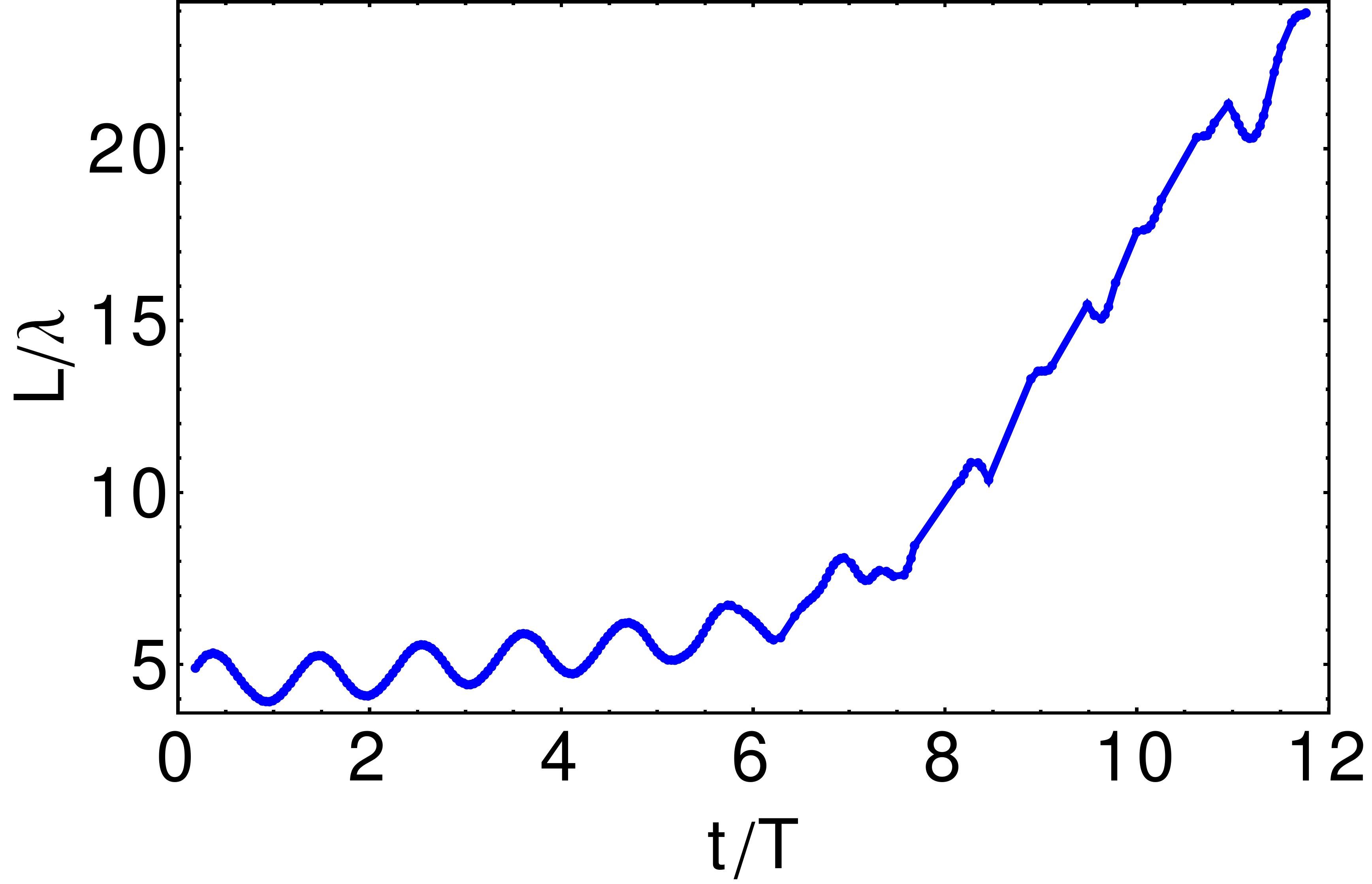}

\caption{Evolution of filament length for a scroll ring propagating in an unbounded
medium and parameter set 1. Parameter values are $\epsilon_{u}=0.07$,
$\epsilon_{w}=\epsilon_{u}/90$, $f=1.4$, $q=0.002$, and $\phi=0.023$.
\label{fig:LengthSRFree}}
\end{figure}

\begin{figure*}
\begin{centering}
\includegraphics[width=1\linewidth]{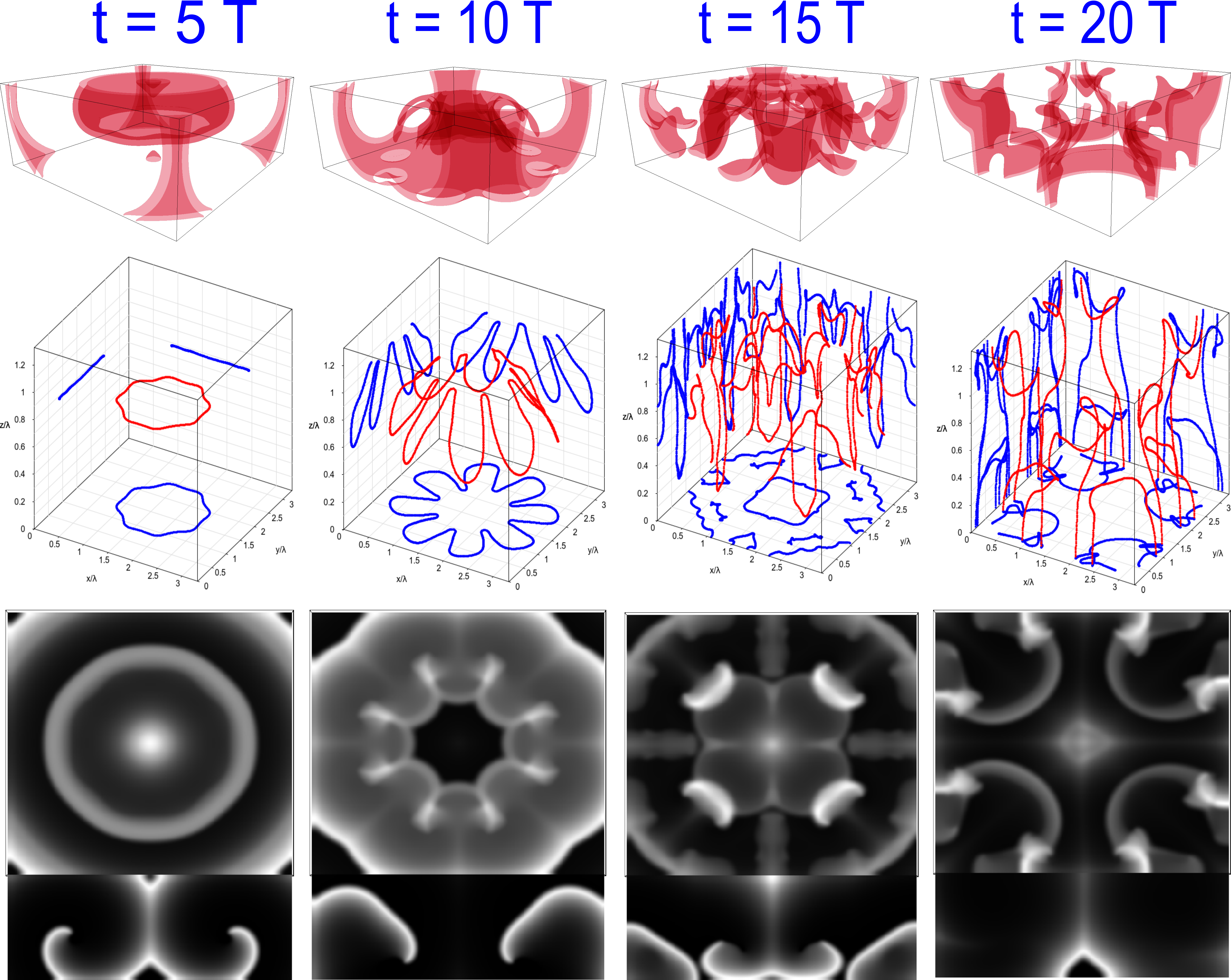}
\par\end{centering}

\caption{Multiple view panels for a scroll ring in an unbounded medium: perspective, 
filament, top, and side view (from top to bottom of figure). Perspective view displays the isosurfaces
for Oregonator variable $u=0.2$, filament view is showing the
filament (red) and its projections into the different planes (blue), while the top view illustrates
the overall concentration of Oregonator variable $v$ as a sum over medium height (as may be seen by an observer 
of an experiment in the PBZR when looking from top). Figure displays destabilization of the filament due to negative filament
tension instability. The initially planar circular filament is distorted
($t=5T$), the modulations increase ($t=10T$), and finally the filament
fragments into pieces ($t=15T$). The cascade results in a spatio-temporally
irregular wave pattern known as Winfree-turbulence. Parameters as
in Fig. \ref{fig:LengthSRFree}.\label{fig:SRFreeMultView}}
\end{figure*}

Initiation of scroll rings in unbounded and bounded media were accomplished
in two steps. First we preparated outwardly propagating cylindric
waves of different radiae as initial conditions while the height of
cylinders were reaching from bottom to top of corresponding media
(from $z=z_{\text{bottom}}\equiv0$ to $z=z_{\text{top}}$). Secondly,
in a cuboidal part of corresponding media the photochemically induced
bromide flow $\phi$ was set to a large value ($\phi=0.2$), in order
to inhibit wave propagation in these regions, starting at simulation
time $t=0$ and resetting $\phi$ homogeneously to standard vaues
(sets 1 and 2) at time $t=1$. After resetting the medium the remained
half cylindric waves could evolve into perfectly planar scroll rings.

Both in unbounded and bounded media the cuboidal region in which the
wave propagation was inhibited was chosen to be the upper part of
the media (see Fig. \ref{fig:SimSpaceConditions}). For unbounded
media this region comprised half of the whole medium while for bounded
media the cuboidal region was chosen such that the initial plane of
corresponding scroll ring filaments were located approximately $0.2\,\lambda$
from the lower Neumann boundary. 

The filament was defined via the crossing of two isosurfaces, namely
$u_{f}=0.3$ and $v_{f}=0.1$ that is the definition of the instantaneous
filament. This definition leads to an oscillation of the filament
position in time.

\section{Scroll ring dynamics in unbounded media\label{sec:Scroll-ring-unbounded}}

Far from any boundary the dynamics of a free scroll ring in homogeneous
media is governed by the equations \cite{Keener:PhysicaD:88,Biktashev:PhilTransRSocA:94}

\begin{align}
\frac{dR}{dt} & =-\frac{\alpha}{R},\label{eq:filament_ode_free_one}\\
\frac{dz}{dt} & =\frac{\beta}{R},\label{eq:filament_ode_free_two}
\end{align}

where $\alpha$ and $\beta$ are constants. Solutions to equations
(\ref{eq:filament_ode_free_one}) and (\ref{eq:filament_ode_free_two})
are given by

\begin{align}
R(t) & =\sqrt{R_{0}^{2}-2\alpha t},\label{eq:fitfilament_R}\\
z(t) & =z_{0}-\frac{\beta}{\alpha}R(t).\label{eq:fitfilament_Z}
\end{align}

\begin{table}[b]
\caption{Calculated values with asymptotic standard errors for filament tension
$\alpha$ and vertical drift coefficient $\beta$ by fitting equations
(\ref{eq:fitfilament_R}) and (\ref{eq:fitfilament_Z}) to the corresponding
filament data from the numerical simulations of scroll rings in unbounded
media.\label{tab:alpha_and_beta}}
\begin{ruledtabular}
\begin{tabular}{ccc}
\multicolumn{1}{c}{\textrm{Parameter set}}&
\multicolumn{1}{c}{\textrm{1}}&
\multicolumn{1}{c}{\textrm{2}}\\
\hline
\\[0.01cm]
$R_{0}\,\left[\lambda\right]$ & $0.70\pm0.01$ & $1.240\pm0.004$ \\[0.3cm]
$\ensuremath{\alpha\,\left[\frac{\lambda^{2}}{T}\right]}$ & 
$\ensuremath{-0.0285\pm0.0014}$ & 
$\ensuremath{-0.0282\pm0.0006}$ \\[0.3cm]
$\beta\,\left[\frac{\lambda^{2}}{T}\right]$ & 
$\ensuremath{0.0030\pm 0.0002}$ & 
$\ensuremath{0.00470\pm 0.00006}$ \\[0.3cm]
\end{tabular}
\end{ruledtabular}
\end{table}

The filament tension $\alpha$ determines whether the scroll ring
shrinks ($\alpha>0$) or expands ($\alpha<0$). A wave with closed
filament and positive line tension would finally collapse while an
open filament would straighten. In both cases small distortions decay.
If the filament tension is negative, a scroll ring will grow. A straight/circular
filament would not remain straight/circular, but small distortions
will grow comparably faster, because of the radius dependence. This
is the negative line tension instability that leads to Winfree turbulence
\cite{Alonso:Science:03a,Zaritski:PhysRevLett:04,Alonso:Chaos:2003}.
The second parameter, $\beta$, leads to a constant drift parallel
to the symmetry axis of the ring. 

We calculated filament tension $\alpha$ and drift coefficient $\beta$
for the chosen parameter sets by fitting equations \ref{eq:fitfilament_R}
and \ref{eq:fitfilament_Z} to the corresponding filament data from
the numerical simulations. Results are shown in table \ref{tab:alpha_and_beta}.

Evolution of the scroll ring is shown in figure \ref{fig:SRFreeMultView}.
The initially circular filament starts to evolve small deviations  that 
grow due to the negative filament tension instability
($t=5T$ to $t=10T$). Finally the filament fragments
into pieces ($t=15T$). Each single fragment grows further
due to the instability ($t=20T$). When each of these single filaments touch
one of the boundaries, they start to fragment further until the wohle
medium is loaded with filament fragments. This is known as scroll
wave (or Winfree) turbulence \cite{Alonso:Science:03a,Zaritski:PhysRevLett:04,Alonso:Chaos:2003}.

\section{Scroll ring interacting with a Neumann boundary\label{sec:Scroll-ring-bounded}}

\begin{figure}
\begin{centering}
\includegraphics[width=1\columnwidth]{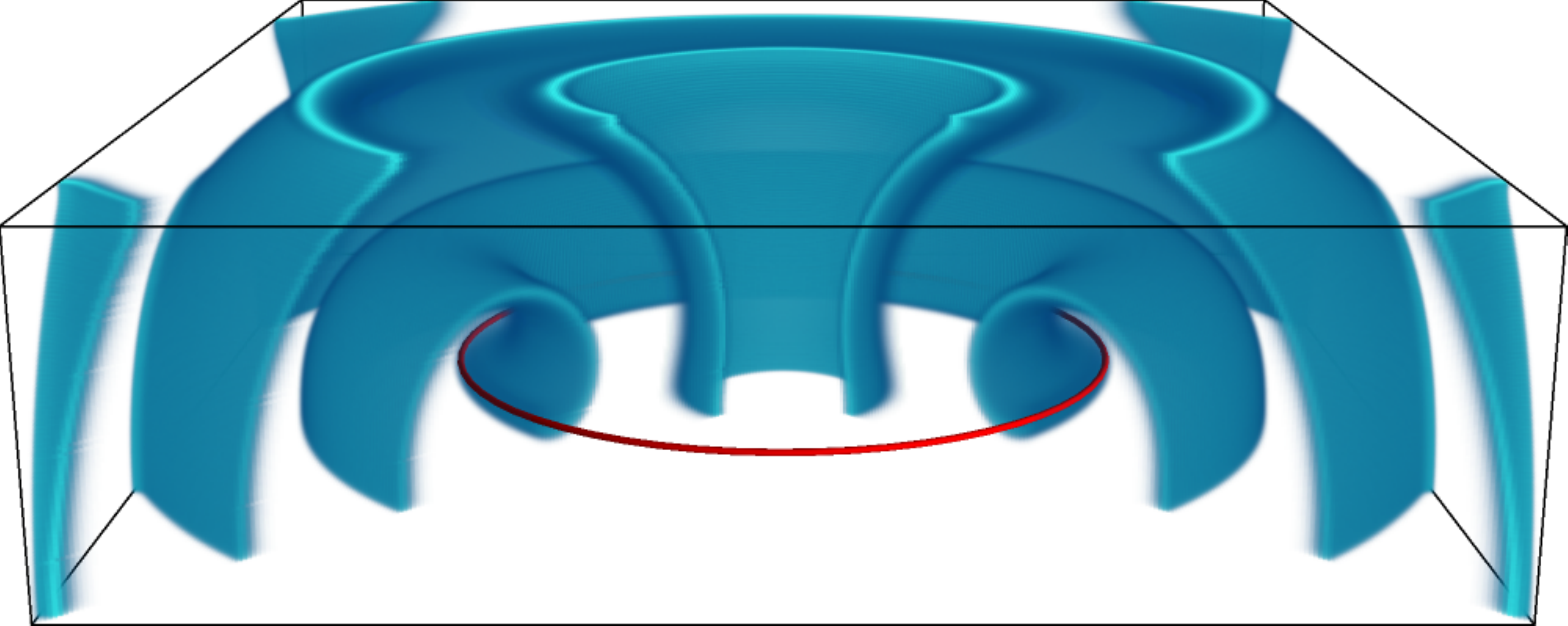}
\par\end{centering}

\caption{Schematic snapshot of a scroll ring within interaction distance to
the lower Neumann boundary. In blue is shown the iso-concentration
planes for one of the model variables and in red the ring-shaped filament.\label{fig:ExampleScrollring}}
\end{figure}

\begin{figure}
\begin{centering}
\includegraphics[width=1\columnwidth]{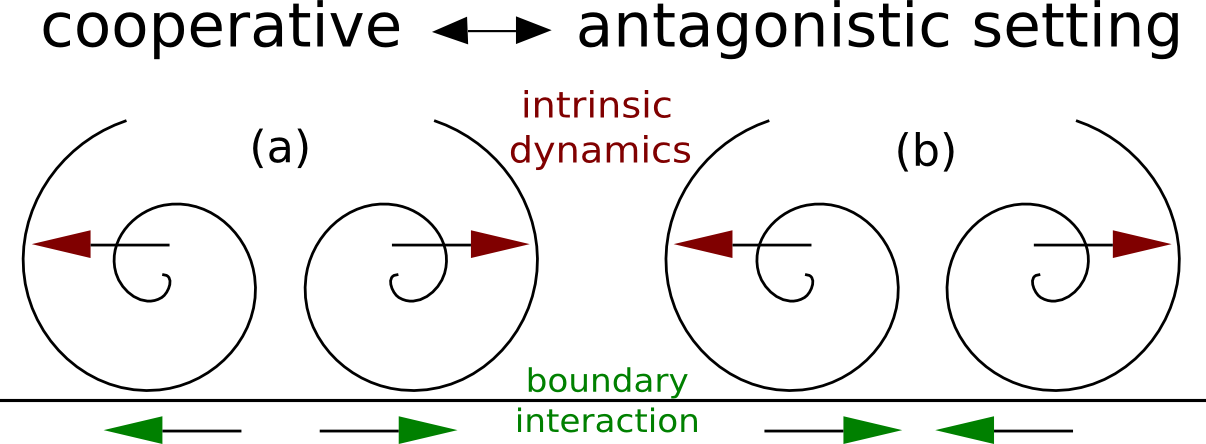}
\par\end{centering}

\caption{Cooperative (a) and antagonistic (b) setting for a scroll ring at a plane
Neumann boundary. Intrinsic and boundary-induced dynamics indicated by full and 
dotted arrows, respectively.\label{fig:SketchCoopAntag}}
\end{figure}

\begin{figure*}
\begin{centering}
\includegraphics[width=1\linewidth]{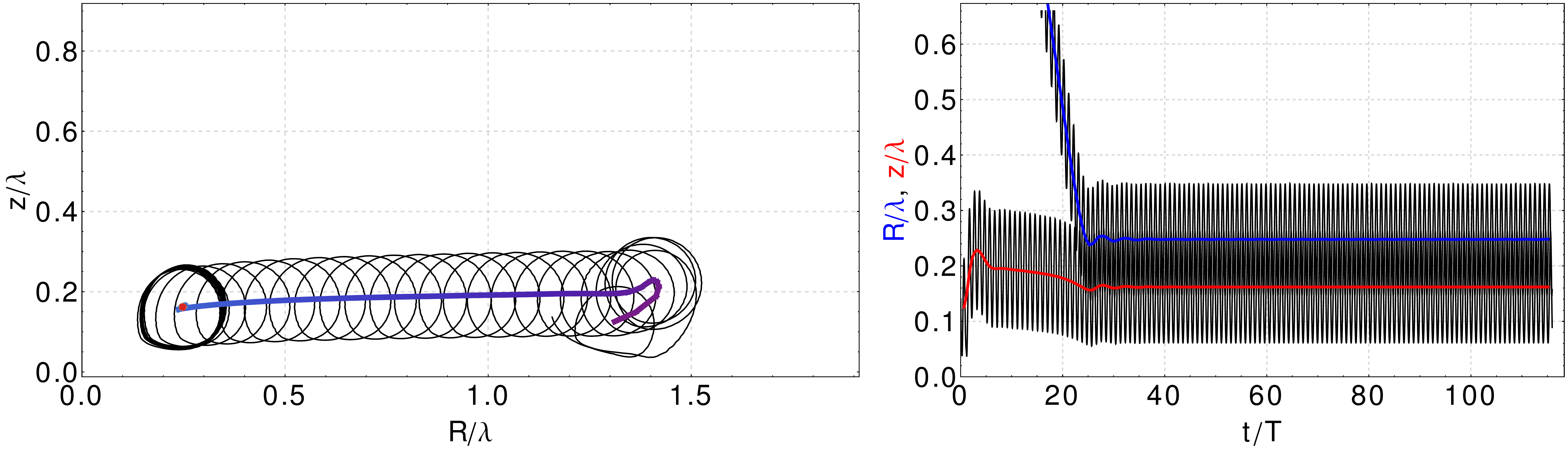}
\par\end{centering}

\caption{Plots for radius $R$ over the $z$- coordinate (left) and $R$ and
$z$ over time $t$ (right) for a scroll ring propagating in a bounded
medium and parameter set 2. The blue and red curves on the right panel
are the calculted means for the radius $\bar{R}(t)$ and mean $z$-coordinate
$\bar{z}(t)$, respectively. Due to the interaction with the boundary
a stable scroll ring of finite radius is formed while the negative
filament tension instability is suppressed. Values for mean radius
and $z$- position are $\bar{R}=0.250\pm0.003\,\lambda$, $\bar{z}=0.155\pm0.003\,\lambda$.
Simulation conducted for parameter set 2. \label{fig:BoundZandRoverTime-Par2}}
\end{figure*}

\begin{figure*}
\begin{centering}
\includegraphics[width=1\linewidth]{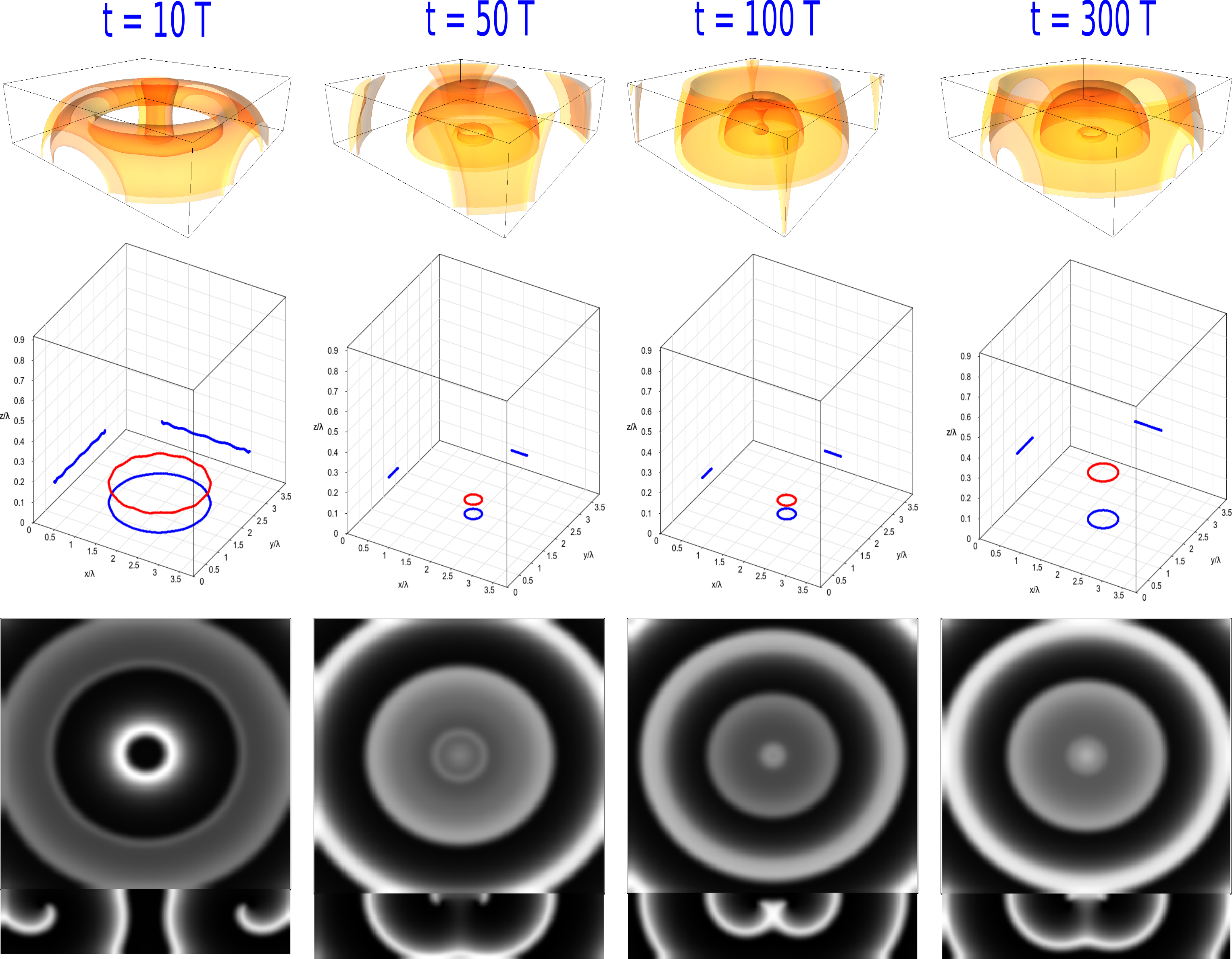}
\par\end{centering}

\caption{Multiple view panels for a scroll ring in a bounded medium: perspective, filament, top, and side view.
(from top to bottom of figure). Perspective view displays the isosurfaces
for Oregonator variable $u=0.2$, filament view is showing the
filament (red) and its projections into the different planes (blue), while the top view illustrates
the overall concentration of Oregonator variable $v$ as a sum over medium height (as may be seen by an observer 
of an experiment in the PBZR when looking from top). Figure displays stabilization of the filament at the boundary. 
Simulation conducted for parameter set 2. \label{fig:SRConfinedMultView}}
\end{figure*}

To investigate how the interaction with a Neumann boundary influences
the dynamics of an expanding scroll ring, we initiated scroll rings
for chosen parameter sets as is illustrated in Fig. \ref{fig:SimSpaceConditions}
for bounded media. \\
Fig. \ref{fig:ExampleScrollring} shows an example
for such a scroll ring that is displaying the isosurfaces of a scroll
ring interacting with the lower Neumann boundary. The distance was chosen to be smaller than the distance at which a
two dimensional spiral starts to interact with a Neumann boundary
for the chosen parameters.

The interaction of a scroll ring with a Neumann boundary can either
increase or decrease its radius. For a scroll ring that is intrinsically
expanding both changes of the radius (due to intrinsic dynamics and
boundary interaction) have the same sign. This setting we will indicate
as cooperative (Fig. \ref{fig:SketchCoopAntag}a), while the second
case, where boundary interaction has the opposite effect on the radius
evolution, we will correspond to as antagonistic setting (Fig. \ref{fig:SketchCoopAntag}b). 

The example scroll ring (Fig. \ref{fig:ExampleScrollring})
interacts with the lower boundary in an antagonistic setting. Such a
scroll ring does not expand but shrink (Fig. \ref{fig:BoundZandRoverTime-Par2})
until it reaches a mean stationary radius $\bar{R}$ and a mean stationary
$z$-position $\bar{z}$ (besides of the oscillation due to the filament
definition). In Fig \ref{fig:SRConfinedMultView} time evolution of the boundary-stabilized
scroll ring is presented by different snapshots and view modes. Obviously the ring
is contracting until it has reached a stable radius und it remains stable up to
$300$ periods of the corresponding spiral wave.

This state is stable for very long times (more than $400$ rotation
periods of corresponding spiral wave). Here we only present the first
$300$ periods for the sake of comprehensibility. For additional data 
(e.g. graphs, videos, etc.) the interested reader is referred to the 
supplementary material to this paper.

The formation of a boundary-stabilized scroll ring can be understood qualitatively if additional terms are introduced
in the kinematic equations for the filament radius and the drift (equations
\ref{eq:filament_ode_free_one} and \ref{eq:filament_ode_free_two},
respectively) that account for the boundary effects (see reference
\cite{Totz:2013})

\begin{align}
\frac{dR}{dt} & =-\frac{\alpha}{R}+c_p\left(z\right)+c_n\left(R\right),\label{eq:filament_ode_bound_01}\\
\frac{dz}{dt} & =\frac{\beta}{R}+c_n\left(z\right)-c_p\left(R\right),\label{eq:filament_ode_bound_02}
\end{align}

with general spiral drift velocity functions $c_p\left(x\right)$ and $c_n\left(x\right)$,
depending on radius $R$ and $z$-position of scroll ring filament, that is $x\in\left \{ z,R \right \}$. 

$c_p\left(z\right)$ could for example strengthen the intrinsic time evolution
of filament (cooperative setting, Fig. \ref{fig:SketchCoopAntag}a)
or weaken/suppress it (antagonistic setting, Fig. \ref{fig:SketchCoopAntag}b).

In order to quantify boundary interaction of scroll rings and find corresponding spiral 
drift velocity fields $c_p$ and $c_n$, we investigated
systematically the interaction of a spiral wave in two spatial dimensions
at a plane Neumann boundary \cite{Totz:2013}. 

\begin{figure}
\begin{centering}
\includegraphics[width=1\columnwidth]{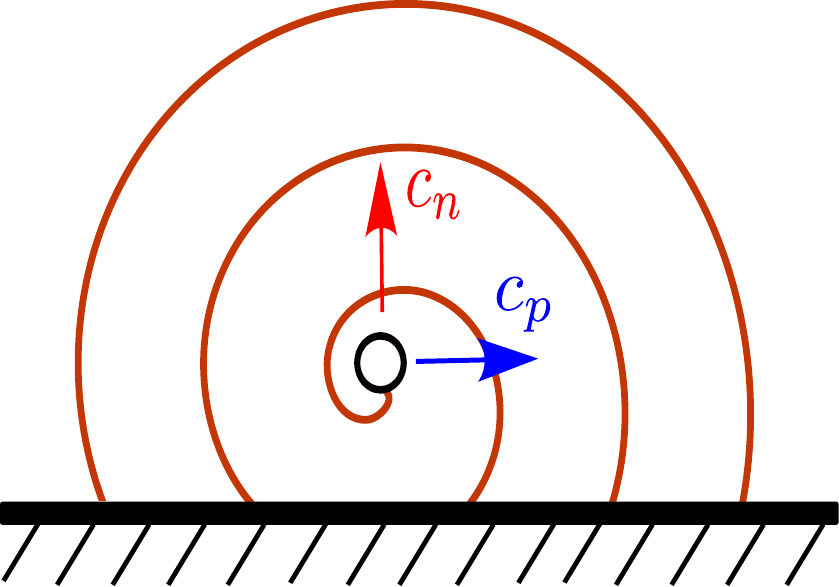}
\par\end{centering}

\caption{Sketch of a spiral at a plane Neumann boundary displaying
boundary-induced drift velocities in parallel $c_{p}$ and normal
direction $c_{n}$ to the boundary.\label{fig:SpiralAtBoundary}}
\end{figure}

We initiated the spiral at different distances to the boundary and
calculated the resulting drift velocities of the corresponding spiral
core in parallel and normal direction to the boundary. This is shown
schematically in Fig. \ref{fig:SpiralAtBoundary}.

\begin{figure}
\begin{centering}
\includegraphics[width=1\columnwidth]{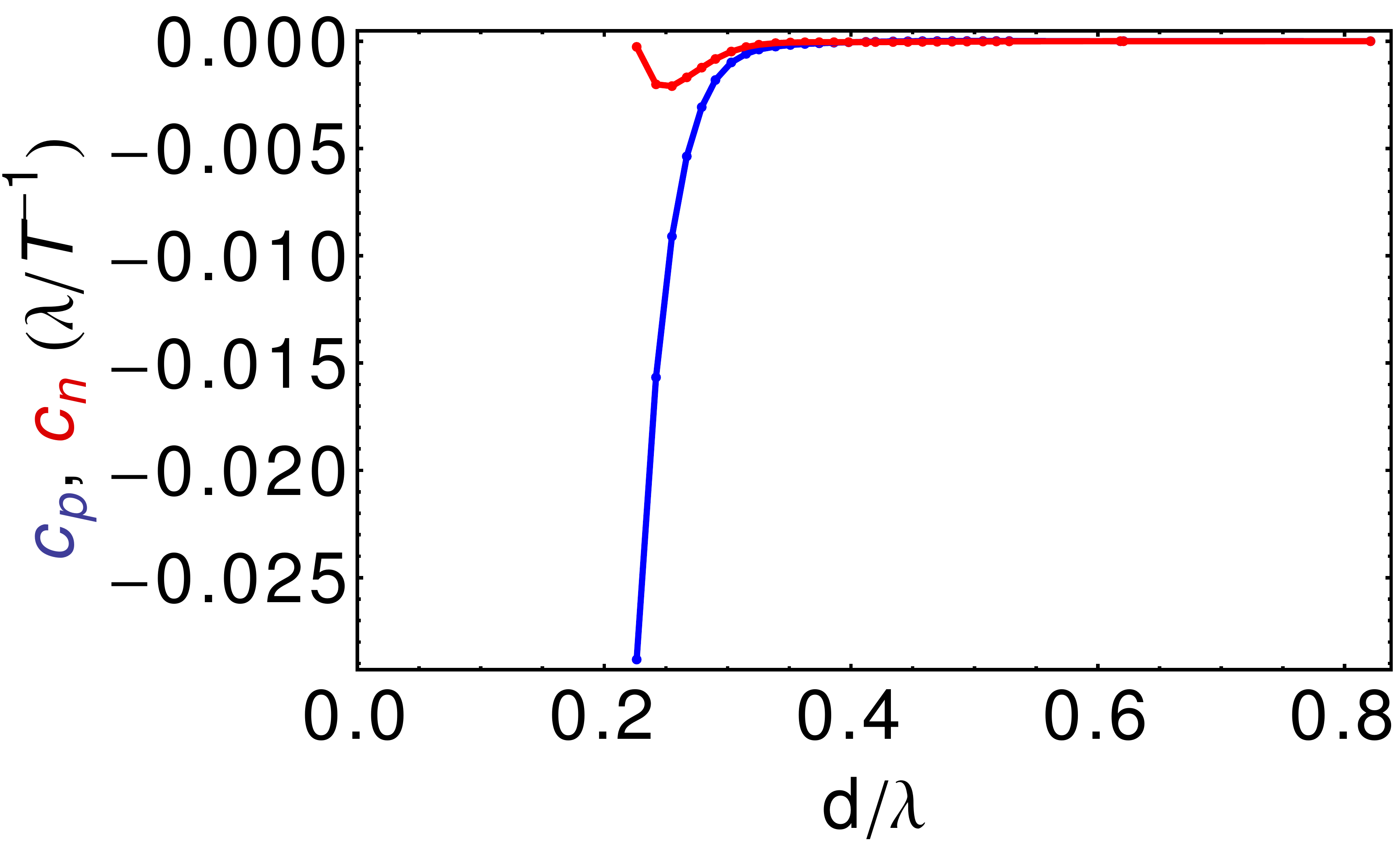}
\par\end{centering}

\caption{Calculated parallel and normal drift velocities, $c_{p}$ and $c_{n}$
respectively, for a spiral initiated at different distances from the
boundary. Simulations conducted for parameter set 2.\label{fig:Measured-cp_and_cn}}
\end{figure}

Results are shown in Fig. \ref{fig:Measured-cp_and_cn}. As one can
see, the absolute value of both drift velocity components are very
small, as long as the spiral core is more than $\sim0.3\,\lambda$ away
from the boundary. As the spiral core reaches regions between
$0.2\,-\,0.3\,\lambda$ away from the boundary, the absolute values
of the two velocity components are strongly increasing. The parallel
drift velocity component reaches a maximum absolute value of $|c_{p}|\approx0.029$.

The calculated value of the scroll ring filament tension for parameter set 2 is 
$\alpha\approx-0.028$ (see table \ref{tab:alpha_and_beta}). Imagine now the scroll 
ring in each radial cross-section to be formed of two counter-rotating
spirals (as is displayed in Fig. \ref{fig:SketchCoopAntag}). Thus,
in each radial cross-section such counter-rotating double-spirals
drift in direction to each other. This is the manifestation of the
antagonistic setting shown in Fig. \ref{fig:SketchCoopAntag}(b).
Since the absolute value for the parallel drift velocity is almost
equal to the absolute value of the ring filament tension, expansion
of the ring is stopped while the ring is stabilized at the boundary.

\section{Formation of boundary-stabilized scroll rings in an exciatble medium - 
conditions for observation in the PBZR \label{sec:Towards_Exp_Verification}}

In the previous section we have shown that an intrinsically expanding
scroll ring can be stabilized at plane Neumann boundaries. This could
be explained phenomenologically by boundary-induced drift of spiral
waves in two spatial dimensions.

But until now we investigated the topic at a more conceptional level
without considering at least the most important deficiencies that
are arising in real experimental systems.

In the next three subsections we will consider some important effects
that come into play when working with photosensitive Belousov-Zhabotinskii
reactions in thin gel layers loaded with silica and the catalyst Ruthenium
\cite{Linde:PhysicaD:1991,Amemiya:PRL77:1996}. Effects that may hinder boundary-induced
stabilization. In subsection \ref{sub:Height_Vs_Lightintensity} we
will examine first at which combinations of medium height and intensity
of photoinhibition stable scroll rings can be initiated at all. Subsection
\ref{sub:Inhomogeneous} will carry on with the question if boundary-induced
stabilization is still possible at differnt degrees of inhomogeneity
in $z$-direction. Finally, in subsection \ref{sub:Inclined-scroll-rings}
scroll rings will be initiated such that the filament plane
will be inclined with refernce to the boundary.

\subsection{Initiation under variation of medium height and intensity of photoinhibition\label{sub:Height_Vs_Lightintensity}}

In experiments with thin layers of photosensitive Belousov-Zhabotinskii
media it is important to check whether boundary-stabilized scroll rings, 
as was presented in the previous section, can be initiated at all.

This is not trivial since two major problems have to be tackled. The
first problem deals with the question for the minimum medium thickness
to support formation of boundary-stabilized scroll rings. The second problem
to solve comes with the photoinhibitory character of the system itself.
As was already shown in references \cite{Amemiya:PRL77:1996,Amemiya:Chaos:98},
scroll rings can be intiated in the photosensitive BZ media by considering
illumination of the medium in the $z$-direction from above or below.
The incident light is attenuated due to absorption in the medium.
Thus, an illumination gradient in $z$-direction will be the consequence.
Considering the photosensitive BZR this light gradient would lead
to a gradient of inhibitor $Br^{-}$ concentration in $z$-direction.
Mathematically this light attenuation have been represented by the
Lambert-Beer relation in the following sense:

\begin{equation}
\phi(z)=\phi_{e}\exp\left(-\alpha z\right),\label{eq:Lambert-Beer}
\end{equation}

when illumination from below is considered. This is the
classic form of the Lambert-Beer relation that was also used in references
\cite{Amemiya:PRL77:1996,Amemiya:Chaos:98}. There $\alpha$ was introduced
as a parameter that ``includes the molar absorption coefficient and
concentration of the reduced catalyst, $Ru\left(bpy\right)_{3}^{2+}$'',
while the amplitude variable $\phi_{e}$ can be seen as ``the quantum
efficiency for the photochemical production of $Br^{-}$''.

\begin{figure}
\begin{centering}
\includegraphics[width=1\columnwidth]{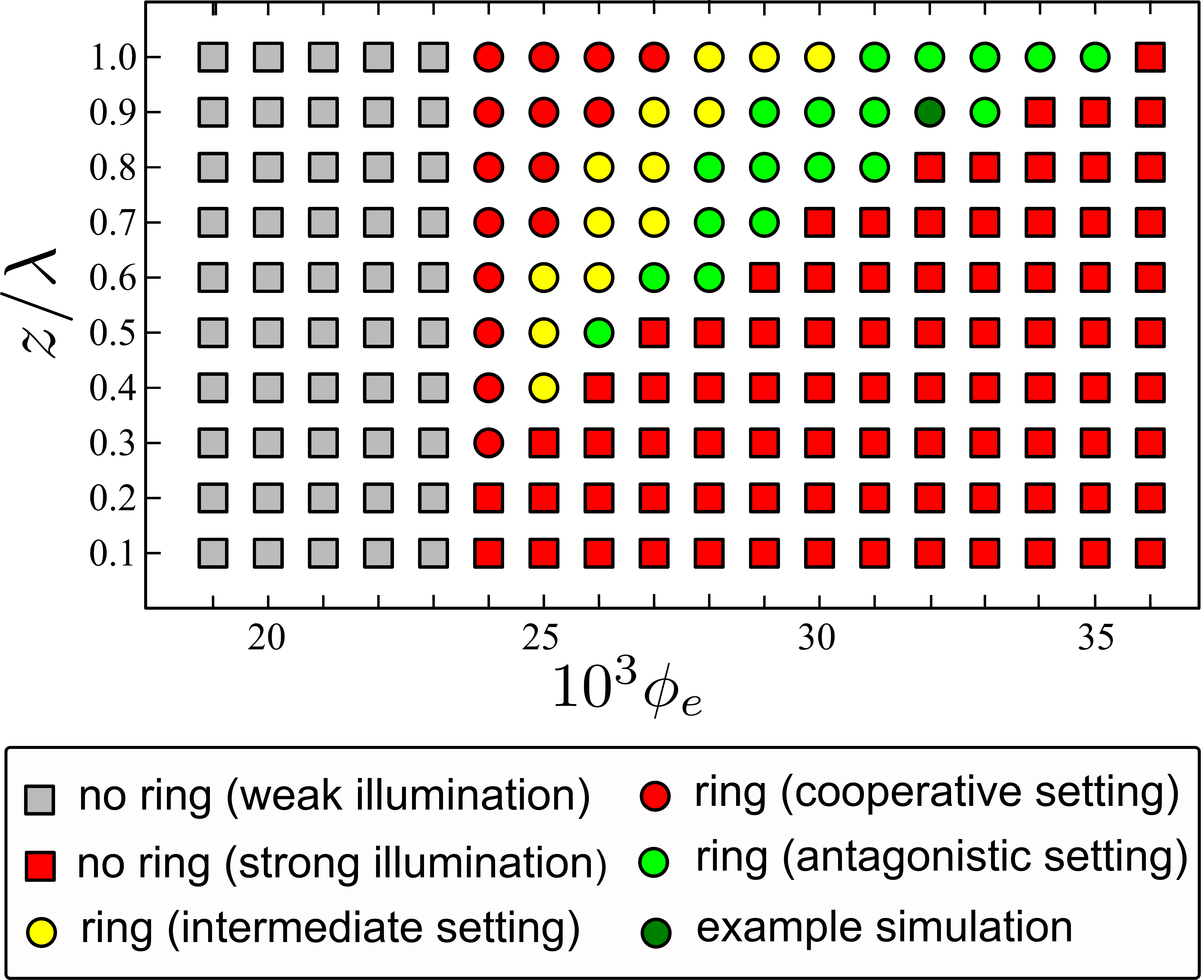}
\par\end{centering}

\caption{Numerical results for scroll ring initiation using partial inhibition
of cylindrical wave fronts by external applied illumination. The photochemically
induced bromide flow $\phi$ (which is assumed to be proportional
to the acting light intensity) is set by the Lambert-Beer relation
(equation \ref{eq:Lambert-Beer}) with $\phi_{0}=0.014$, $\alpha=0.05$,
while medium thickness and the illumination intensity parameter $\phi_{e}$
are varied.\label{fig:Phasediagram}}
\end{figure}

In this work we fixed parameter $\alpha$ to the value $\alpha=0.05$,
while systematically varying thickness of media for chosen values
of the illumination intensity parameter $\phi_{e}$, in order to check
for which value pairs of $z$ and $\phi_{e}$ a stable scroll ring
can be initiated. The high-intensity illumination that is represented
by equation \ref{eq:Lambert-Beer} will be turned on at time $t_{\text{start}}=0\, t.u.$
and turned off at time $t_{\text{end}}=2\, t.u.$ ($t.u.$ indicates
simulation time units).

\begin{figure*}[t]
\begin{centering}
\includegraphics[width=1\linewidth]{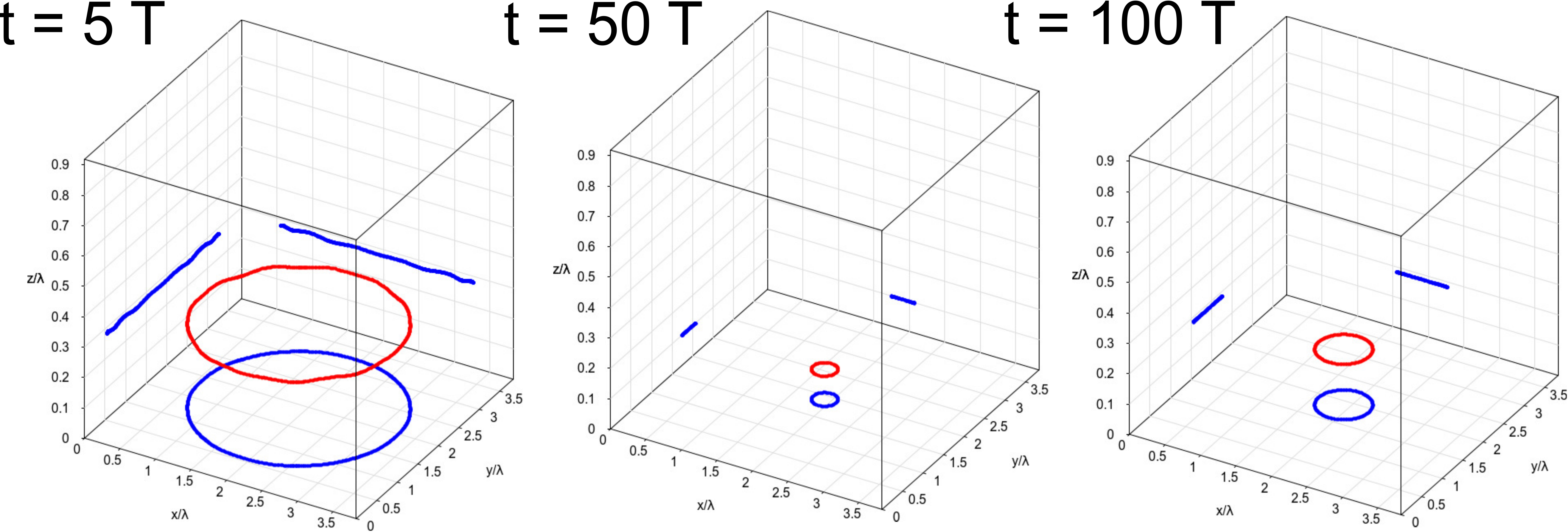}
\par\end{centering}

\caption{Evolution of the filament in the course of time for scroll rings in
a gradient of the illumination parameter $\phi$. Gradients accomplished
by application of the Labert-Beer law $\phi\left(z\right)=\phi_{0}e^{-\alpha z}$
with $\phi_{0}=0.014$ and $\alpha=0.0033$.\label{fig:FilamentGradient_Par02}}
\end{figure*}

The results are shown in Fig. \ref{fig:Phasediagram}. In the complete
region below $10^{3}\phi_{e}=24$ the overall illumination $\phi(z)$
was too weak for initiating stabily rotating scroll rings. This is
indicated by grey squares in the figure. For media with height values
below $0.3\,\lambda$ ($\lambda$ again is indicating the wave length
of the corresponding spiral waves) the media are too thin to support
formation of scroll rings, since illumination intensity values higher
than $10^{3}\phi_{e}=24$ lead to complete inhibition of waves (indicated
by red squares). Red circles represent formation of scroll rings that
are interacting with one of the plane Neumann boundaries in a cooperative
sense (see Fig. \ref{fig:SketchCoopAntag}a). This would lead to an
amplified expansion of the ring. Yellow circles denote scroll rings
initiated such that their filament plane is located in the middle
between the two boundaries. Such initiated rings may vertically drift
to the stabilizing boundary (in the antagonistic setting shown in Fig.
\ref{fig:SketchCoopAntag}b). Finally, by green circles the truely
boundary-stabilized scroll rings are represented. In these cases,
the layer thickness as well as the illumination intensity are appropriate
for initiating scroll rings that are from the start perfectly placed
in the interaction regime to the stabilizing boundary. A combination,
by which the resulting ring would be analogous to the example case
from previous section, is also shown here by a dark green circle.

\subsection{Excitability gradient in parallel to scroll ring's symmetry axis \label{sub:Inhomogeneous}}

In the previous subsection we demonstrated how formation of boundary-stabilized
scroll rings could be achieved in the framework of the photosensitive
Belousov-Zhabotinskii reaction. This was done by setting up a temporarily
inhomogeneous photoinhibition in $z$-direction based upon the Lambert-Beer
relation, while the further time evolution of the scroll ring was
accomplished under homogeneous illumination. 

For testing stability of a boundary-induced pacemaker in a more realistic
experimental situation one has to take into account spatially inhomogeneous
illumination during the wole simulation time. 

Here we took a very early state (after $2$ rotation periods) of the
scroll ring for parameter set 2, presented in section
\ref{sec:Scroll-ring-bounded}, as initial state for futher investigation
in an illumination gradient with two different gradient strengths.
This was achieved by setting the light parameter $\phi$ inhomogeneously
through the Lambert-Beer relation,
\[
\phi\left(z\right)=\phi_{0}\exp\left(-\alpha z\right)
\]
from below, $\phi_{0}=0.014$, and two different absorption parameters, (1) $\alpha_{1}=0.0033$,
and (2) $\alpha_{2}=0.01$.

Exemplarily, evolution of filament for absorption parameter 1
is shown in Fig. \ref{fig:FilamentGradient_Par02}. For both gradient
strengths the scroll ring is stable more than $400$ rotation periods
of the corresponding spiral wave. Here, for reasons of comprehensibility,
evolution for the first $100$ periods is shown. As one can see also
the shape of the scroll ring filament does not change much. This means
that the scroll ring at the no-flux boundary acts as a periodic source of waves,
at least over a relative long time.

\subsection{Inclined initiation\label{sub:Inclined-scroll-rings}}

\begin{figure*}
\begin{centering}
\includegraphics[width=1\linewidth]{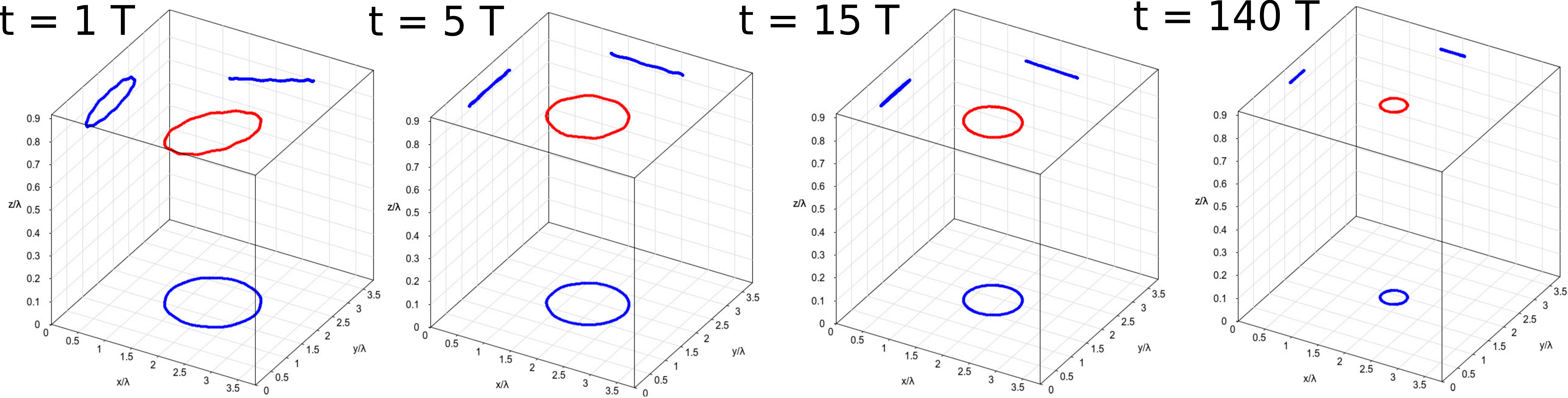}
\par\end{centering}

\caption{Evolution of an inclined scroll ring filament up to $140$ rotation
periods of the corresponding spiral wave. Different snapshots of the
filament (red) with its projections on the sidewalls (blue) are shown.
Initiation by application of the function \ref{eq:inclining_function}
with $\phi_{e}=0.05$, $\alpha=0.1$, $\phi_{i}=0.01$, and $L=125$.
Simulation carried out for parameter set 2. \label{fig:FilamentInclined_Par01}}
\end{figure*}

Under some circumstances initiation of scroll rings in experiments
for lightsensitive BZR may also happen inhomogeneously in horizontal
($x-$ and/or $y$-) direction, such that scroll rings can not be
initiated perfectly planar but in some way inclined. 

Here we want to adress the question if a boundary-induced stabilization
nevertheless can be reached when a scroll ring is initiated in an
inclined fashion, such that the initial phase of boundary interaction
is not equal for different points of the corresponding ring-shaped
filament.

To investigate inclined scroll rings numerically, we have chosen cylindric-shaped
waves as initial conditions, like it was done in the previous sections
for planar scroll rings (see Fig. \ref{fig:SimSpaceConditions}),
but this time we intiated the inclined scroll rings by setting the
Oregonator model parameter $\phi=\phi\left(x,z\right)$ according
to the following function
\begin{equation}
\phi\left(x,z\right)=\phi_{0}+\phi_{e}\exp\left(-\alpha z\right)+\frac{\phi_{i}}{L}x,\label{eq:inclining_function}
\end{equation}

turned on at time $t_{\text{start}}=0$ and turned off at time $t_{\text{end}}=0.5$.
Here the second term, characterized by parameters $\phi_{e}$ and
$\alpha$, is the well-known Lambert-Beer relation. These two parameters
will be fixed, $\phi_{e}=0.05$ and $\alpha=0.1$. The last term introduces
inclination of the final scroll ring. The degree and strength of inclination
can be controlled by parameters $\phi_{i}$, and $L$.

We conducted simulations for scroll rings at the antagonistic boundary
with two different inclination states: $\phi_{i}=0.01$ and (1) $L=125$
and (2) $L=100$. For case 1 the inclination angle relative to the boundary
at time $t=1T$ had a value of $\gamma _\text{inc} \approx 4^\circ$, while for 
inclination state 2 this value was $\gamma _\text{inc} \approx 5^\circ$. 
In both cases the scroll ring stabilized, up to $300$ rotation periods of corresponding
spiral wave. Exemplarily, this is shown for the inclination state 1
in Fig. \ref{fig:FilamentInclined_Par01}.  As one can observe
here, the primarily inclined ring-shaped filament reaches a stable planar
state already after $5$ periods.

\section{Conclusion\label{sec:Conclusion}}

We have shown that it is possible to stabilize an intrinsically expanding
scroll ring via the interaction with a Neumann boundary. The negative
line tension instability and Winfree-turbulence are suppressed, and instead the 
ring is stabilized. Finally, conditions were discussed under 
which scroll rings with negative filament tension could be stabilized in 
thin layers of the PBZR.

\section{Acknowledgements}

We cordially thank the German Science Foundation (DFG) for financial support through the
special research field 910 (sfb 910) and the Research Training Group 1558 (GRK 1558).
Simulations in this paper were accomplished within the interactive reaction-diffusion
simulation environment 
\href{https://savannah.nongnu.org/projects/virtuallslab/}{\textbf{VirtualLab}} that was developed in our work group.
Thus we thank the people (S. Fruhner, F. Paul, S. Molnos, D. Kulawiak, etc.) who helped us to build up this outstanding tool.

\end{document}